\documentclass[journal]{IEEEtran}

\usepackage{amsmath,amssymb}
\usepackage{graphicx}
\usepackage{float}
\usepackage{textcomp}
\usepackage{colortbl}
\usepackage{multirow}
\usepackage{enumitem}
\usepackage[dvipsnames]{xcolor}
\usepackage{tikz}
\usetikzlibrary{positioning}
\usepackage{pgfplots}
\pgfplotsset{compat=1.18}
\usepgfplotslibrary{groupplots}
\PassOptionsToPackage{dvipsnames}{xcolor} 
\usepackage{xcolor}

\graphicspath{{./}{./figures/}} 
\DeclareGraphicsExtensions{.pdf,.png,.jpg}

\usepackage{cite} 

\usepackage{tikz}

\usetikzlibrary{positioning,calc, arrows.meta,shapes.geometric} 

\usepackage{pgfplots}
\pgfplotsset{compat=1.18}

\usepackage{booktabs}

\usepackage{hyperref} 
\hypersetup{
    colorlinks=true, 
    linktoc=all,     
    linkcolor=blue,  
    citecolor=blue,
    urlcolor=blue
}

\def\del#1{\textcolor{Gray}{}}   

\definecolor{lightblue}{RGB}{173,216,230} 

\title{LLM2Manim: Pedagogy-Aware AI Generation of STEM Animations}

\author{%
Aastha Joshi, Hongyi Ke, Meet Gajjar, Aaron Christian, Qi Wang, Jun Chen 
\thanks{Department of Aerospace Engineering, San Diego State University, San Diego, CA 92182 USA (email: qwang4@sdsu.edu).}%
}

\begin{document}

\maketitle

\begin{abstract}

High-quality STEM animations can be useful for learning, but they are still not common in daily teaching, mostly because they take time and special skills to make. In this paper, we present a semi-automated, human-in-the-loop (HITL) pipeline that uses a large language model (LLM) to help convert math and physics concepts into narrated animations with the Python library \texttt{Manim}. The pipeline also tries to follow multimedia learning ideas like segmentation, signaling, and dual coding, so the narration and the visuals are more aligned.

To keep the outputs stable, we use constrained prompt templates, a symbol ledger to keep symbols consistent, and we regenerate only the parts that have errors. We also include expert review before the final rendering, because sometimes the generated code or explanation is not fully correct.

We tested the approach with 100 undergraduate students in a within-subject A-B study. Each student learned two similar STEM topics, one with the LLM-generated animations and one with PowerPoint slides. In general, the animation-based instruction gives slightly better post-test scores (83\% vs.\ 78\%, $p < .001$), and students show higher learning gains ($d=0.67$). They also report higher engagement ($d=0.94$) and lower cognitive load ($d=0.41$). Students finished the tasks faster, and many of them said they prefer the animated format. Overall, these results suggest LLM-assisted animation can make STEM content creation easier, and it may be a practical option for more classrooms.

\end{abstract}

\section{Introduction}
Learning can be more effective when the content is animated and narrated in an intuitive way \cite{zull2023brain,stein2020multisensory}. But in many STEM classes, the material is still shown as static symbols and equations, to an extent analogous to reading sheet music, while you never hear the true music. Multimedia learning research suggests that adding visuals and sounds together (animation with narration) can help students understand more, since the information is not coming from only one form \cite{mayer2002animation,mayer2014cambridge}.

However, making good STEM animations was notoriously difficult for educators. It often is time-consuming and requires coding skills to be combined with teaching experience \cite{hoffler2007instructional}. Manim library facilitates this process, but still requires manual writing to adjust every scene. This makes the process quite demanding and inaccessible for most instructors \cite{hoffler2007instructional,markovic2024manim}.

In this study, we build an LLM + Manim pipeline to make this process more accessible for educators. The system starts from a short natural-language brief, e.g., an instructor's objective teaching a math/physics topic, and then generates a narrated \texttt{Manim} animation for classroom use. This framework is not fully automatic, and still keeps a human-in-the-loop (HITL) check before release, to compensate for possible flaws in the LLM-generated drafts.

\begin{figure*}[t]
  \centering
  \includegraphics[width=\textwidth]{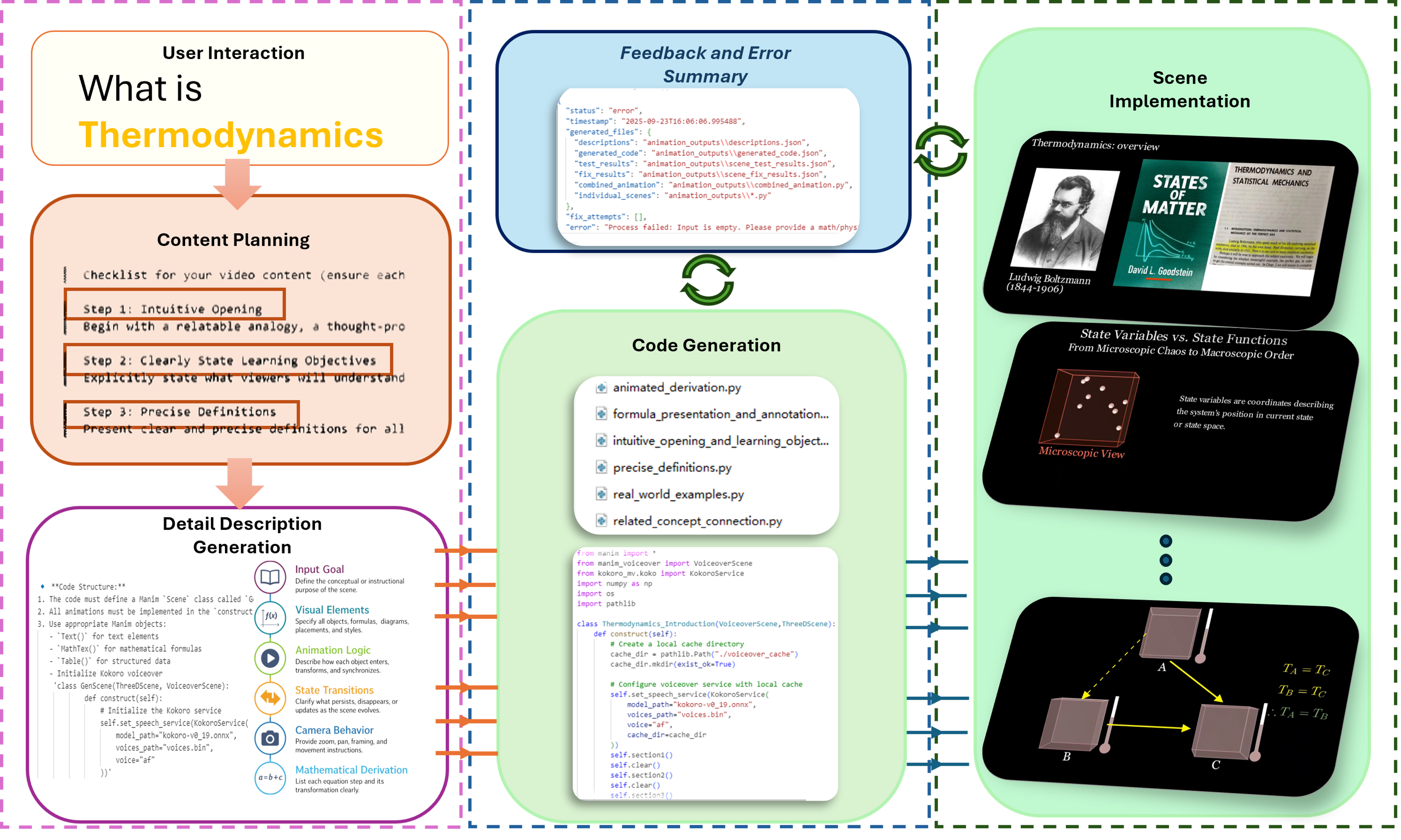}
  \caption{Overview of the LLM-driven, pedagogy-aware animation pipeline. 
  The system converts an instructor's goal or a student's question into a narrated STEM animation through five stages:
  (1) inputs and optional resources; (2) pedagogy-guided scene planning and narration outline;
  (3) automatic Manim code generation with self-correction; (4) rendering with narration, visual synchronization; and
  (5) delivery via an accessible, analytics-enabled platform.}
  \label{fig:system-overview}
\end{figure*}

The workflow is organized into five stages: (i) inputs and resources; (ii) simple scene planning; (iii) \texttt{Manim} code generation with some self-correction; (iv) rendering with synchronized narration; and (v) delivery on a platform with basic analytics. Following such a schematic, a high-level concept can be turned into a full explainer video (Fig.~\ref{fig:system-overview}). The guidance of the workflow also includes common multimedia learning principles, such as segmentation and signaling, making the generated videos easier to follow, although the quality still depends on the topic and the generated output.

\subsection{Background and Related Work}\label{sec:background}

Multimedia learning research often finds learning improvement when verbal and visual information are both included and aligned, instead of only reading text \cite{mayer2014cambridge,sweller2011cognitive}. The Cognitive Theory of Multimedia Learning (CTML) describes two main channels (audio and visual) of the learning process, each with a limited capacity. Without the material being paced, combined, and intuitively organized, students may feel overloaded.

Based on this understanding, some common design rules are used in STEM explanations. For example, spoken narration with graphics is usually better than putting excessive text on the screen. It also helps when speech and visuals happen at the same time (temporal contiguity). Segmentation and signaling, in addition, help to break a long idea into smaller parts and tell students what to focus on \cite{mayer2014cambridge,ploetzner2017strategies,dekoning2017guiding}.

Some meta-analysis studies further suggest that animation can be better than static figures, but mostly when the design is clean and not distracting \cite{berney2016animation,hoffler2007instructional}. If the motion is not aligned, or if there are too many extra details, it backfires to increase cognitive load and reduce learning \cite{mayer2014cambridge,dekoning2009attention}. In the current work, we try to keep narration and visuals more aligned using prompts, templates, and simple checks to minimize the risk of such drawbacks.

Another related view is cognitive load theory (CLT), which focuses on pacing and reducing unnecessary load \cite{sweller2011cognitive}. In math and physics, students often need to track symbols and intermediate steps. If the explanation jumps too rapidly, or the screen has overloaded contents, it confuses the students. In practice, we keep notation stable, align narration with what is shown, and maintain a minimum essential amount of elements at the same time \cite{jain2014aisle}. These simple principles lead to choices that may help keep the outputs more stable and easier to follow.

Visualization tools have also been used historically in STEM education, from early algorithm animations to more modern interactive tools \cite{hubscher2001strategy,miller2006oop}. In many cases, dynamic visuals help students understand processes that are difficult to grasp by static diagrams, since the steps are shown more clearly \cite{lin2010dynamic}.
More recently, Manim makes it possible to create high-quality animations with reasonable math typesetting. Some studies and reports demonstrate its usefulness for explanations, despite the high authoring cost \cite{chavan2024parsing,markovic2024manim}. This is the main motivation to use LLM support to facilitate the instructor to start from a short brief and get a draft quickly, instead of developing everything from scratch \cite{lu2024teachersai}.

\begin{table*}[t]
\caption{Comparison of Visualization/Animation Approaches for STEM Learning and How Our System Advances the State of the Art}
\label{tab:compare}
\centering
\scriptsize
\renewcommand{\arraystretch}{0.95}
\begin{tabular}{p{5cm}ccccc}
\toprule
\textbf{Approach / System} & \textbf{Automation} & \textbf{Pedagogy} & \textbf{Narrative Voice Sync} & \textbf{Evaluations} & \textbf{Reference}\\
\midrule
Manual Manim workflows & $\times$ & $\triangle$ & $\triangle$ & $\triangle$ & \mbox{\cite{chavan2024parsing,markovic2024manim}} \\
\rowcolor{lightblue!30}3D interactive OOP tool & $\triangle$ & $\triangle$ & $\times$ & $\triangle$ & \mbox{\cite{miller2006oop}} \\
Dynamic 3D math graphs & $\times$ & $\triangle$ & $\times$ & $\checkmark$ & \mbox{\cite{lin2010dynamic}} \\
\rowcolor{lightblue!30}Parsing via Python{+}Manim & $\times$ & $\triangle$ & $\triangle$ & $\triangle$ & \mbox{\cite{chavan2024parsing}} \\
Manim for algorithms \& DS & $\times$ & $\triangle$ & $\triangle$ & $\triangle$ & \mbox{\cite{markovic2024manim}} \\
\rowcolor{lightblue!30}Interactive visual learning (ML) & $\times$ & $\checkmark$ & $\times$ & $\checkmark$ & \mbox{\cite{liu2025interactive}} \\
Visualization morphing & $\times$ & $\checkmark$ & $\times$ & $\triangle$ & \mbox{\cite{ruchikachorn2016learning}} \\
\rowcolor{lightblue!30}AI: Manimator & $\checkmark$ & $\times$ & $\triangle$ & $\times$ & \cite{sathvik2025manimator} \\
AI: TheoremExplainAgent & $\checkmark$ & $\triangle$ & $\triangle$ & $\times$ & \mbox{\cite{ku2025theoremexplainagent}} \\
\rowcolor{lightblue!30}Manual Manim (3Brown1Blue) & $\times$ & $\checkmark$ & $\checkmark$ & $\checkmark$ & ~\mbox{\cite{sanderson2025threeb1b}} \\
Code2Video  & $\checkmark$ & $\triangle$ & $\times$ & $\times$ & ~\mbox{\cite{zheng2025code2video}} \\
\rowcolor{lightblue!30}\textbf{Current study: LLM Manim pipeline} & $\checkmark$ & $\checkmark$ & $\checkmark$ & $\checkmark$ & - \\
\bottomrule
\end{tabular}
\begin{flushleft}
\footnotesize Legend: $\checkmark$ present; $\triangle$ partial/implicit; $\times$ absent.
\end{flushleft}
\end{table*}

\subsection{AI-Generated Educational Content and Animation}

The emergence and rapid growth of LLM have made it technically possible to realize automatic generation of narrated, animated teaching material, including simple tasks like lesson outlines, quizzes, and short explanations \cite{lu2024teachersai} \cite{wang2024flippedprompt}
. This technology saves time and helps teachers prepare faster \cite{yin2024chatbotfeedback}.
On the other hand, at the current stage, human manual labor is still required for high-quality animations (for example, from the YouTube channel ``3Brown1Blue"). We foresee that it would still take a period of time until LLMs can fully take over this process to scale up \cite{sanderson2025threeb1b}. Meanwhile, research efforts continuously explore directions towards this ultimate goal. Automatic systems like Code2Video focus on layout automation and code-to-video generation \cite{zheng2025code2video}. Table~\ref{tab:compare} gives a simple summary of these different directions, including manual expert work, code-based systems, and interactive platforms. We compare them on automation level, pedagogy support, narration-visual alignment, and whether they are tested with learners \cite{xu2023semanticnavigation}.

Some recent prototype systems also try to turn technical text into visuals, or generate step-by-step explanations. For example, Manimator converts passages from papers or math descriptions into Manim animations \cite{sathvik2025manimator}. It tries to keep the content faithful, but it does not incorporate strong teaching design rules or classroom validation. TheoremExplainAgent focuses on theorem understanding with multimodal explanations in a narrow math setting \cite{ku2025theoremexplainagent}, but does not report learner results. These works suggest the direction is promising, but it still remains an open question how much students really learn from these generated materials \cite{neyem2024knowledgeassistant}.

In the current work, we aim to fill this gap in a more practical manner. Firstly, we treat narration and visuals as two outputs that should match each other, and we guide them with CTML/CLT ideas. Secondly, we adopt a simple symbol ledger and validators to keep notation and units stable to prevent potential conflict of symbols or small mistakes. Thirdly, we keep a human-in-the-loop (HITL) review and evaluate the animations with students in a counterbalanced classroom study.
We also note that interactive systems, where learners can change parameters and see results in real time, can be very helpful for exploration \cite{liu2025interactive}. But they usually need custom development and more effort, so we do not include them in the current study.

Overall, our framework aims to keep the generation of automations fast and scalable, but also keeps core pedagogy rules and a structured review before release. As shown in Table~\ref{tab:compare}, this combination may be a practical option for real teaching use.

\subsection{Research Questions and Hypotheses}

This study focuses on five questions based on our design goal.
\begin{enumerate}[label={\textbf{RQ\arabic*}:}, left=0.2em]
    \item Do narration-synchronized, LLM-generated animations lead to higher post-test scores and larger learning gains than matched slides?
    \item Does the animation condition increase engagement while keeping the workload similar or lower?
    \item Are the benefits similar or maybe larger for students with lower prior knowledge or other underserved subgroups?
    \item Does animation help students finish concept tasks faster under limited class time?
    \item Do students report higher satisfaction with animation-based materials?
\end{enumerate}

Based on these questions, we expected better learning performance (\textbf{H1}) and higher engagement without higher workload (\textbf{H2}). We also expected the effect may be more helpful for lower-prepared learners (\textbf{H3}), with shorter task time (\textbf{H4}), and higher satisfaction (\textbf{H5}). 
For analysis, we mostly follow common practice. We use Analysis of covariance (ANCOVA) for post-tests with pre-test as a covariate. For within-subject outcomes, we use paired tests, and we report effect sizes and confidence intervals.

The remainder of this paper is organized as follows. Section \ref{sec:methodology} introduces the proposed authoring pipeline and its narration-visual alignment and reliability mechanisms. Section \ref{sec:experiments} outlines the classroom study design, participants, instruments, and analysis procedures. Section \ref{sec:results} reports the results. Sections \ref{sec:discussion} and \ref{sec:conclusion} conclude with a discussion of the findings, equity considerations, limitations, and future directions.

\section{Methodology}
\label{sec:methodology}

\begin{figure}[!t]
  \centering
  \includegraphics[width=\linewidth]{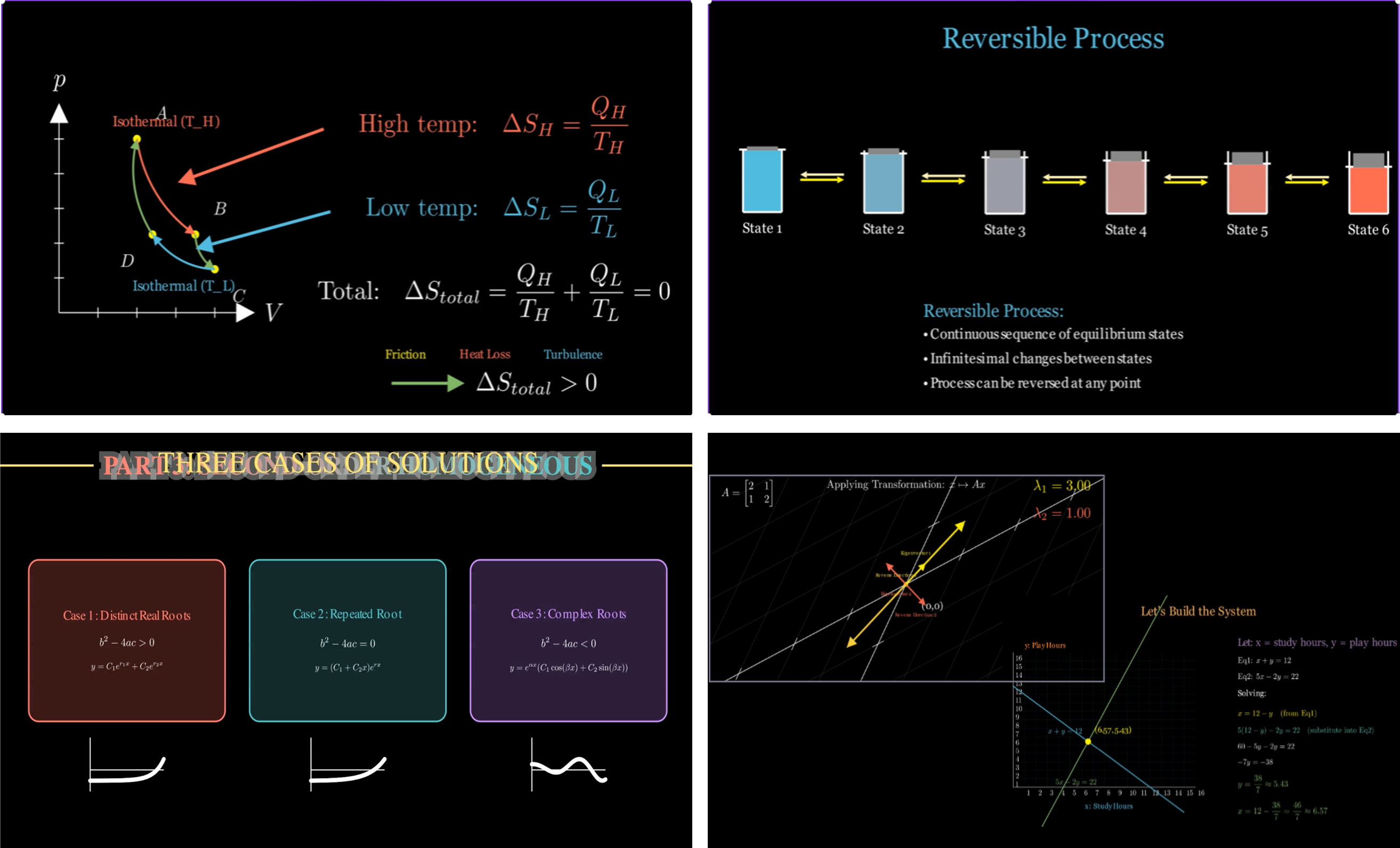}
  \caption{Example outputs from the pipeline. The bottom scenes show a common layout issue (text and figure overlap), so we fix it by a quick human edit.A representative rendered animation is available at:
\url{https://youtu.be/cUnw-wGVlUk}.}
  \label{fig:sample_output}
\end{figure}

\begin{figure*}[htbp]
    \centering
    \includegraphics[width=\textwidth]{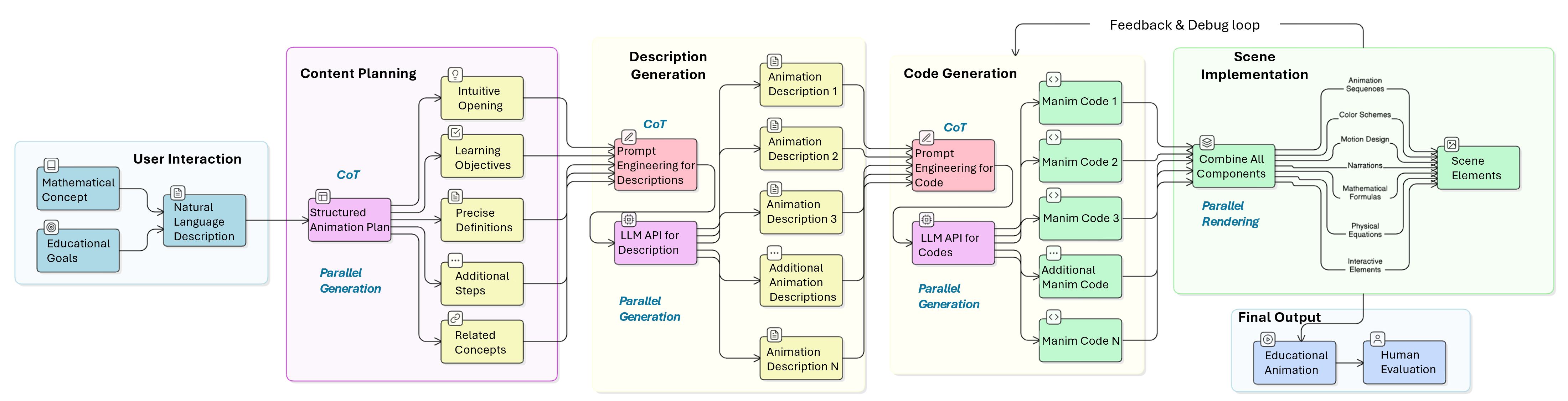}
    \caption{
        \label{fig:structure graph}
        Overall pipeline of the HITL authoring system, from a short brief to a rendered video.
    }
\end{figure*}

We build a semi-automated pipeline that turns a math/physics topic into a short teaching video (3-10 minutes). Fig.~\ref{fig:sample_output} shows typical outputs. It also shows a common problem, including label overlaps or spacing issues, which is further adjusted with HITL. 
The full workflow is:
\emph{User Interaction} $\rightarrow$ \emph{Content Planning} $\rightarrow$ \emph{Parallel Processing} $\rightarrow$ \emph{Scene Assembly} $\rightarrow$ \emph{Final Output}.
The schematic is shown in Fig.~\ref{fig:structure graph}, with a detailed explanation in the following sections.

\subsection{Plan First, Then Generate}

\begin{figure}[H]
\centering
\begin{tikzpicture}[
    font=\small,
    node distance=6pt,
    slot/.style={
        draw,
        rounded corners,
        align=center,
        minimum width=0.80\linewidth,
        inner sep=5pt
    }
]
\node[slot] (brief) {\textbf{Concept Brief}};
\node[slot, below=of brief] (objectives) {\textbf{Scene Goals}};
\node[slot, below=of objectives] (ledger) {
    \textbf{Symbol List}\\
    {\footnotesize notation, units, assumptions}
};
\node[slot, below=of ledger] (narration) {
    \textbf{Narration Cues}\\
    {\footnotesize short timed segments}
};
\node[slot, below=of narration] (storyboard) {\textbf{Storyboard Frames}};
\node[slot, below=of storyboard] (codehints) {
    \textbf{Code Constraints}\\
    {\footnotesize layout, timing, allowed primitives}
};
\node[slot, below=of codehints] (validation) {
    \textbf{Checks}\\
    {\footnotesize units, symbol consistency}
};
\draw[very thick, rounded corners]
($(brief.north west) + (-0.35,0.35)$)
rectangle
($(validation.south east) + (0.35,-0.35)$);
\end{tikzpicture}

\caption{
Slot-based plan template to make key items explicit. Once an issue is identified, instead of revisiting everything, only the broken part will be modified and fixed.
}
\label{fig:prompt-structure}
\end{figure}
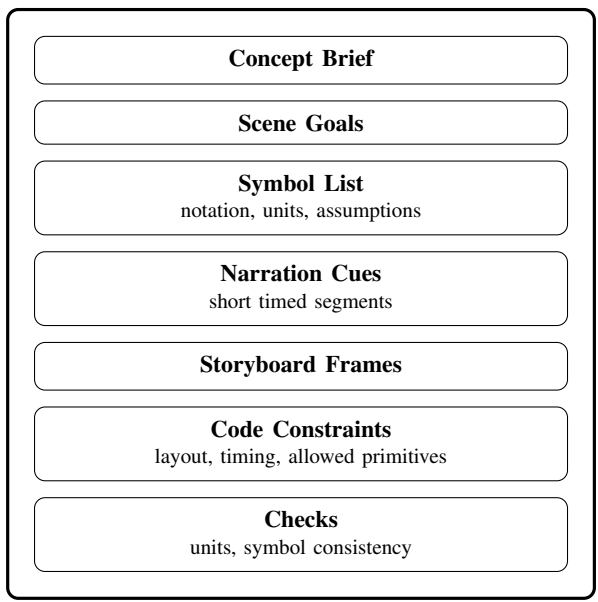

The process starts with a short concept brief to create alignment with the students' background knowledge and define the concept to explain. A structured plan is built afterwards to help keep outputs stable across scenes.
The plan includes a few simple parts: (1) scene goals, (2) a symbol list with units and assumptions, (3) short narration cues, (4) storyboard frames, and (5) code constraints (layout and timing), as shown in Fig.~\ref{fig:prompt-structure} . Through experiments, such planning demonstrates much better stability with reduced small drift, symbol changes, and tangential divergence of concepts within the video.

\subsection{Alignment between Speech and Visuals}

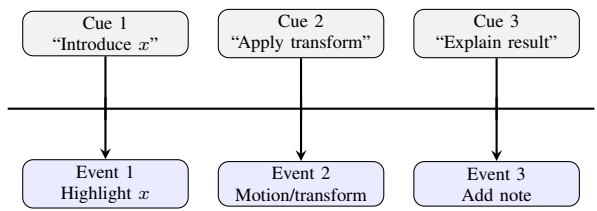
\begin{figure}[H]
\centering
\begin{tikzpicture}[
    font=\scriptsize,
    event/.style={draw, rounded corners, minimum height=0.55cm, inner sep=2pt, align=center},
    >=stealth
]
\draw[thick] (0,0) -- (7.8,0);

\node[event, fill=gray!10, minimum width=2.2cm] (n1) at (1.3,1.0) {Cue 1\\``Introduce $x$''};
\node[event, fill=gray!10, minimum width=2.2cm] (n2) at (3.9,1.0) {Cue 2\\``Apply transform''};
\node[event, fill=gray!10, minimum width=2.2cm] (n3) at (6.5,1.0) {Cue 3\\``Explain result''};

\node[event, fill=blue!8, minimum width=2.2cm] (v1) at (1.3,-1.0) {Event 1\\Highlight $x$};
\node[event, fill=blue!8, minimum width=2.2cm] (v2) at (3.9,-1.0) {Event 2\\Motion/transform};
\node[event, fill=blue!8, minimum width=2.2cm] (v3) at (6.5,-1.0) {Event 3\\Add note};

\draw[->, thick] (n1.south) -- (v1.north);
\draw[->, thick] (n2.south) -- (v2.north);
\draw[->, thick] (n3.south) -- (v3.north);

\foreach \x in {1.3, 3.9, 6.5}{
    \draw[thick] (\x,0.15) -- (\x,-0.15);
}
\end{tikzpicture}

\caption{
Speech-visual alignment. Cues are linked to events to keep the videos easier to follow.
}
\label{fig:synchronization-timeline}
\end{figure}

A prominent difficulty in the current work is to make speech and visuals align in time (\emph{temporal contiguity}).
To solve such a difficulty, we applied simple timing markers. Narration cues are linked to visual events. If they drift, either the timing will be adjusted or the whole part will be regenerated. Fig.~\ref{fig:synchronization-timeline} shows a simple example for this technique.

In practice, a topic is also split into small parts (\emph{segmentation}). Most scenes are 60-120 seconds. This makes the pace steadier and easier to debug.

\subsection{Parallel Drafting and Merging}

As illustrated in Fig.~\ref{fig:parallel-generation}, the narration and Manim code are generated in parallel for a faster response in real time. This further avoids one error spreading to everything, if narration is fine but code breaks, only the code will be regenerated for that scene and vice versa.

\begin{figure}[H]
\centering
\begin{tikzpicture}[
    font=\scriptsize,
    node distance=6pt,
    box/.style={
        draw, rounded corners, align=center,
        minimum width=2.5cm, minimum height=0.7cm, inner sep=3pt
    }
]
\node[box, fill=gray!10] (np1) {Prompt};
\node[box, fill=gray!10, below=of np1] (np2) {Cues};
\node[box, fill=gray!10, below=of np2] (np3) {Storyboard};
\node[box, fill=gray!10, below=of np3] (np4) {Symbol list};

\node[box, fill=blue!8] (cp1) [right=1.7cm of np1] {Code prompt};
\node[box, fill=blue!8, below=of cp1] (cp2) {Primitives};
\node[box, fill=blue!8, below=of cp2] (cp3) {Layout hints};
\node[box, fill=blue!8, below=of cp3] (cp4) {Timing marks};

\node[box, fill=green!10, below=1.0cm of np4, minimum width=5.5cm]
(merge) {Merge narration \& code};

\draw[->, thick] (np1) -- (np2);
\draw[->, thick] (np2) -- (np3);
\draw[->, thick] (np3) -- (np4);

\draw[->, thick] (cp1) -- (cp2);
\draw[->, thick] (cp2) -- (cp3);
\draw[->, thick] (cp3) -- (cp4);

\draw[->, thick] (np4.south) -- (merge.north -| np4.south);
\draw[->, thick] (cp4.south) -- ++(0,-0.4) -| (merge.north);
\end{tikzpicture}

\caption{
Parallel generation. Two tracks are separate, then merged. This makes fixes cheaper when only one part is wrong.
}
\label{fig:parallel-generation}
\end{figure}
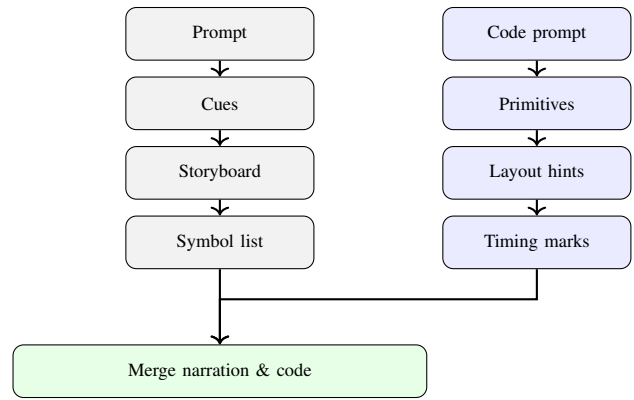




\begin{figure}[H]
\centering
\begin{tikzpicture}[
    font=\scriptsize,
    node distance=6pt,
    box/.style={draw, rounded corners, align=center,
        minimum width=6.3cm, minimum height=0.75cm, inner sep=3pt},
    >=stealth
]
\node[box, fill=gray!10] (syn) {Run check\\{\footnotesize imports, \LaTeX{}, deterministic run}};
\node[box, fill=gray!10, below=of syn] (bind) {Cue/event check\\{\footnotesize timing alignment}};
\node[box, fill=gray!10, below=of bind] (ledger) {Symbol/unit check\\{\footnotesize keep meaning stable}};
\node[box, fill=gray!10, below=of ledger] (ped) {Goal coverage\\{\footnotesize show the key step}};
\node[box, fill=green!10, below=of ped] (merge) {Merge};
\node[box, fill=blue!8, below=of merge] (render) {Render or regenerate part};

\draw[->, thick] (syn) -- (bind);
\draw[->, thick] (bind) -- (ledger);
\draw[->, thick] (ledger) -- (ped);
\draw[->, thick] (ped) -- (merge);
\draw[->, thick] (merge) -- (render);
\end{tikzpicture}

\caption{
Checks before rendering. We keep them simple and mostly local, because many errors are small and easy to fix.
}
\label{fig:scene-assembly}
\end{figure}
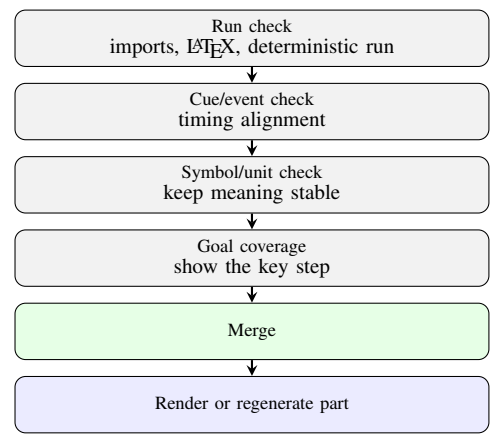

\subsection{Human Review (HITL)}

As mentioned in the introduction, the current work is not fully automated and still requires human review. Three quick-pass criteria are adopted: subject-matter, teaching quality, and engineering. 
LLMs often drift in notation and style. Code can also break after library updates. To mitigate these issues, we adopt several practical safeguards, including structured prompting, low-temperature decoding for code generation, restricted primitive operations, and a curated set of regression test scenes. When templates or dependencies change, we regenerate the regression scenes and compare them with older outputs. Significant deviations are then identified and examined to ensure consistency and correctness.
A build manifest (model id, prompt version, seeds, Manim/\LaTeX{} versions) is also stored along the way to help with reproducibility and debugging. Before final rendering, a set of lightweight validation checks is applied to each scene to ensure correctness and alignment, as summarized in Fig.~\ref{fig:scene-assembly}.

In the current study, we focus on short videos (3-10 minutes) as a proof of concept for a long-term goal of automatic generation of course contents. The review cost stays reasonable so far. Longer videos are possible, but require more fixes and stability. Interactive variants are explored but not yet included in the classroom study in this paper.

\section{Experiments}
\label{sec:experiments}
The study involves $N = 100$ students enrolled in mathematics, physics, aerospace engineering, computer science, and information systems courses at San Diego State University. Participant demographic characteristics are summarized in Table~\ref{tab:participant-demographics}.  The sample represents typical STEM majors aged 18-26 years ($M = 21.4$, $SD = 2.3$) with varying prior knowledge levels (15\% none, 35\% basic, 35\% intermediate, 15\% advanced). Participation is voluntary and uncompensated for grades; students receive a \$10 digital gift card for each completed survey as appreciation.
\begin{table}[H]
\centering
\renewcommand{\arraystretch}{1.2}
\caption{Participant Demographic Characteristics in the A-B Crossover Study (N = 100).}
\label{tab:participant-demographics}
\begin{tabular}{p{0.38\linewidth} p{0.52\linewidth}}
\hline
\textbf{Category} & \textbf{Value} \\
\hline
Age (Mean $\pm$ SD) & 22.29 $\pm$ 2.48 \\
Age Range & 18-26 years \\
Majors & Mathematics (26\%), Computer Science (24\%),\\
& Physics (20\%), Aerospace (15\%), Information Systems (15\%) \\
Prior Knowledge & Intermediate (42.5\%), Basic (36.2\%),\\
& Advanced (15.0\%), None (6.3\%) \\
\hline
\end{tabular}
\end{table}

The study was approved by the university's Institutional Review Board, and informed consent was obtained from all participants prior to data collection. The study was conducted under standard educational research ethics protocols. Data were anonymized prior to analysis, and participants could withdraw at any time without penalty.

\subsection{Research Design and Procedures}
A within-subjects A-B crossover design was employed to minimize inter-individual variability. Each participant completed two parallel learning modules, one delivered through LLM-generated Manim animations and the other via traditional PowerPoint slides, covering comparable topics (Linear Transformations, Linear Systems, Eigenvalues and Eigenvectors, Thermodynamics). To control order effects, participants were randomly assigned to one of two sequences: Animation-First followed by Slides-Second, or Slides-First followed by Animation-Second. 

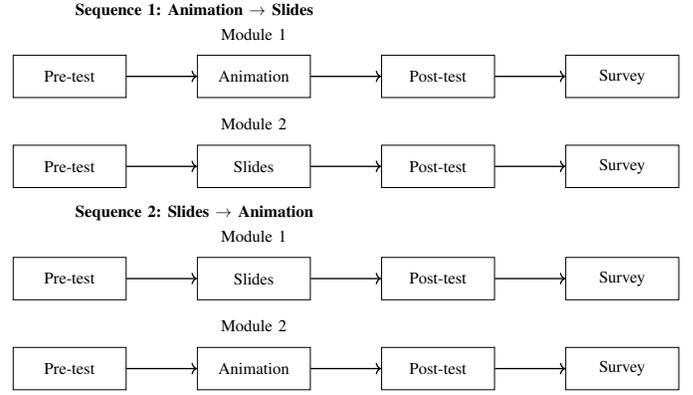
\begin{figure}[t]
\centering
\resizebox{\columnwidth}{!}{%
\begin{tikzpicture}[
    box/.style={draw, rectangle, minimum width=2.4cm, minimum height=0.9cm, align=center},
    arrow/.style={->, thick},
    node distance=1cm and 1.5cm
]
\node[anchor=west] at (0,3.4) {\textbf{Sequence 1: Animation $\rightarrow$ Slides}};

\node (s1m1pre) [box] at (0,2) {Pre-test};
\node (s1m1anim) [box, right=of s1m1pre] {Animation};
\node (s1m1post) [box, right=of s1m1anim] {Post-test};
\node (s1m1sur) [box, right=of s1m1post] {Survey};
\draw[arrow] (s1m1pre) -- (s1m1anim);
\draw[arrow] (s1m1anim) -- (s1m1post);
\draw[arrow] (s1m1post) -- (s1m1sur);
\node[above=0.2cm of s1m1anim] {Module 1};

\node (s1m2pre) [box, below=of s1m1pre] {Pre-test};
\node (s1m2slide) [box, right=of s1m2pre] {Slides};
\node (s1m2post) [box, right=of s1m2slide] {Post-test};
\node (s1m2sur) [box, right=of s1m2post] {Survey};
\draw[arrow] (s1m2pre) -- (s1m2slide);
\draw[arrow] (s1m2slide) -- (s1m2post);
\draw[arrow] (s1m2post) -- (s1m2sur);
\node[above=0.2cm of s1m2slide] {Module 2};

\node[anchor=west] at (0,-0.9) {\textbf{Sequence 2: Slides $\rightarrow$ Animation}};

\node (s2m1pre) [box] at (0,-2.3) {Pre-test};
\node (s2m1slide) [box, right=of s2m1pre] {Slides};
\node (s2m1post) [box, right=of s2m1slide] {Post-test};
\node (s2m1sur) [box, right=of s2m1post] {Survey};
\draw[arrow] (s2m1pre) -- (s2m1slide);
\draw[arrow] (s2m1slide) -- (s2m1post);
\draw[arrow] (s2m1post) -- (s2m1sur);
\node[above=0.2cm of s2m1slide] {Module 1};

\node (s2m2pre) [box, below=of s2m1pre] {Pre-test};
\node (s2m2anim) [box, right=of s2m2pre] {Animation};
\node (s2m2post) [box, right=of s2m2anim] {Post-test};
\node (s2m2sur) [box, right=of s2m2post] {Survey};
\draw[arrow] (s2m2pre) -- (s2m2anim);
\draw[arrow] (s2m2anim) -- (s2m2post);
\draw[arrow] (s2m2post) -- (s2m2sur);
\node[above=0.2cm of s2m2anim] {Module 2};

\end{tikzpicture}%
}
\caption{Within-subject A-B crossover design with counterbalanced order. 
Sequence~1 completed the Animation module first, followed by Slides, while 
Sequence~2 completed Slides first, followed by Animation. Each module included 
a pre-test, instructional phase, post-test, and survey, with distinct topics for 
Module~1 and Module~2.}
\label{fig:crossover_design}
\end{figure}

Each module followed a uniform timeline consisting of a pre-test (5-10 min) to assess baseline understanding, an instructional phase (15 min) for studying the assigned format individually, a post-test (5-10 min) to measure learning gains, and a perception survey (5 min) capturing engagement and workload responses along with open-ended reflections. All content was pedagogically identical in text and visuals, differing only in delivery medium. The same instructor developed both versions and was blinded to conditions during grading to prevent bias. Because the study employed a within-subjects A-B crossover design, module order was counterbalanced to control for sequence and carryover effects. The impact of order (Animation-First vs. Slides-First) was later examined as a between-subjects factor to verify that observed learning, engagement, and workload differences were not attributable to ordering artifacts.

\subsection{Data Collection}
Learning outcomes were measured using matched pre- and post-quizzes, consisting of five expert-validated conceptual and procedural questions per topic, targeting application-level learning outcomes. Engagement was measured with the Intrinsic Motivation Inventory (IMI), a six-item, seven-point Likert scale (1 = Strongly Disagree to 7 = Strongly Agree) assessing enjoyment, value, and interest (e.g., ``I enjoyed learning with [animation/slides]''). The IMI items and anchors are reproduced in the survey instruments. Cognitive workload was measured using the NASA-TLX instrument with six subscales (Mental, Physical, Temporal Demand, Performance, Effort, and Frustration), each rated on a 0-20 scale (0 = Very Low to 20 = Very High), following the standardized format shown in the survey forms. Participants also provided open-ended feedback on clarity, pacing, and overall learning experience. Responses were coded thematically by two independent raters, and intercoder reliability exceeded Cohen's $\kappa = 0.80$. In this study, $\kappa$ was used to measure agreement between raters while accounting for chance agreement. Cohen’s $\kappa$ is defined as
\begin{equation}
\kappa = \frac{p_o - p_e}{1 - p_e},
\end{equation}
where $p_o$ is the observed proportion of agreement and $p_e$ is the expected agreement by chance \cite{cohen1960kappa}.

\subsection{Data Analysis Plan}
To support interpretation of the reported results, we report both reliability and effect size metrics alongside inferential statistics. Internal consistency of multi-item scales was assessed using Cronbach’s $\alpha$, which indicates how consistently the survey items measure the same intended concept (e.g., engagement or cognitive workload); values above 0.70 are generally considered acceptable. Cronbach’s $\alpha$ is defined as
\begin{equation}
\alpha = \frac{k}{k-1}\left(1 - \frac{\sum_{i=1}^{k} \sigma_i^2}{\sigma_T^2}\right),
\end{equation}
where $k$ is the number of items, $\sigma_i^2$ is the variance of item $i$, and $\sigma_T^2$ is the variance of the total score \cite{cronbach1951alpha}. For within-subject comparisons, effect sizes are reported using Cohen’s $d$ for paired samples, which describes how large the difference between instructional conditions is, independent of sample size, with values of 0.2, 0.5, and 0.8 corresponding to small, medium, and large effects. For paired samples, Cohen’s $d$ is defined as
\begin{equation}
d = \frac{\bar{X}_1 - \bar{X}_2}{s_d},
\end{equation}
where $\bar{X}_1 - \bar{X}_2$ represents the mean difference between conditions and $s_d$ is the standard deviation of the paired differences \cite{cohen1988statistical}. Together, these metrics help clarify both the reliability of the measures and the practical importance of the observed differences.

To ensure direct alignment between the research questions and the analytical procedures, each RQ was mapped to a corresponding statistical test and measurement strategy. 

\textbf{RQ1 (Learning Achievement).} This question was evaluated through two complementary analyses:  
(a) an ANCOVA, with post-test score as the dependent variable, instructional condition as the fixed factor, and pre-test score as a covariate, enabling an adjusted comparison of learning performance; and  
(b) paired-samples \textit{t}-tests on learning gains (post-pre) to examine within-student differences across conditions.

\textbf{RQ2 (Engagement and Cognitive Load).} This question was addressed using paired-samples \textit{t}-tests applied to the IMI engagement scale and the NASA-TLX workload index, respectively. These tests allowed us to quantify whether students reported higher motivation or reduced perceived effort when learning with animations.

\textbf{RQ3 (Equity and Accessibility).} This question was explored through subgroup analyses based on prior-knowledge levels and demographic attributes, examining whether the animation condition provided disproportionate benefits to lower-prepared or underserved learners.

All analyses incorporated appropriate effect sizes (partial $\eta^{2}$ for ANCOVA and Cohen’s $d$ for paired tests) and 95\% confidence intervals to support interpretation of practical significance.

All quantitative analyses were conducted in Python using NumPy, SciPy, and StatsModels. Inferential tests included an ANCOVA with post-test as dependent variable, instructional condition as factor, and pre-test as covariate ($post = \beta_0 + \beta_1 condition + \beta_2 pre + \varepsilon$). Partial $\eta^2$ was computed to quantify effect size for the ANCOVA.
Paired-samples $t$-tests compared Animation and Slides conditions for learning gains, engagement, and workload. Effect sizes were computed using Cohen’s d for paired samples, with 95\% confidence intervals. Significance threshold was set at $\alpha = 0.05$ (two-tailed). Qualitative responses were analyzed inductively, and emergent themes were triangulated with quantitative trends. All data, code, and figure scripts are archived for reproducibility.

To provide a clearer sense of measurement stability, we also calculated 95\% confidence intervals for the reliability estimates using a bootstrap resampling procedure. For the IMI engagement scale, Cronbach's $\alpha$ was $0.82$ with a confidence interval of [$0.78$, $0.86$]. The NASA-TLX workload scale showed $\alpha = 0.79$ with an interval of [$0.74$, $0.83$]. These intervals indicate that both instruments performed consistently in our sample and that the observed internal consistency is unlikely to be due to sampling variability.

\subsection{Satisfaction Analysis}
To address RQ5, overall learner satisfaction with each instructional medium was measured using a single Likert-type item included in both perception surveys. Participants rated their satisfaction on a 1-7 scale immediately after completing each module. Because the study used a within-subjects crossover design, satisfaction differences between the Animation and Slides conditions were analyzed using paired-samples \textit{t}-tests. Cohen’s \textit{d} and 95\% confidence intervals were computed to quantify the magnitude and precision of the observed effect.

Satisfaction was treated as a complementary affective measure distinct from engagement, capturing holistic learner preference and perceived value of the instructional experience. This analysis provides insight into the acceptability and experiential quality of the generated animations, supporting evaluation of the system beyond learning and workload outcomes.

\subsection{Equity and Subgroup Analysis}

RQ3 investigated whether the animation condition provided disproportionate benefits for learners with lower prior knowledge or other underserved subgroups. To examine this question, we conducted subgroup analyses based on participants’ self-reported prior-knowledge levels (None, Basic, Intermediate, Advanced). For statistical power and interpretability, these levels were collapsed into two groups: Low (None + Basic) and High (Intermediate + Advanced). 

Participant-level learning gain differences (Animation - Slides) were computed by pivoting the dataset to wide format. Independent-samples \textit{t}-tests compared gain differences across prior-knowledge groups, and complementary ANCOVA models tested for condition~$\times$~subgroup interactions while controlling for baseline pre-test performance. These analyses evaluated whether animations offered additional scaffolding that particularly supported lower-prepared learners.

Additional robustness checks assessed whether the observed learning advantage varied by demographic factors (e.g., major, age), instructional sequence (Animation-First vs.\ Slides-First), topic, or module period. Order and topic effects were examined using mixed models and regression terms. These analyses ensured that the benefits of the animation condition were not attributable to sequencing artifacts or topic-specific difficulty, strengthening the internal validity of the experimental findings.

\subsection{Validity and Ethics}
Internal validity was ensured through within-subject counterbalancing, identical content, and blinded grading. External validity was supported by recruiting students from diverse STEM majors, enhancing generalizability. Construct validity was upheld through the use of validated IMI and NASA-TLX instruments and expert-verified quiz items. Statistical conclusion validity was supported by testing assumptions of normality and homogeneity, with no significant violations observed; outliers beyond three interquartile ranges were inspected and retained. Ethical compliance included IRB approval, informed consent, voluntary participation, anonymity, and adherence to IEEE ethics standards for human-subject learning research.

\section{Results}
\label{sec:results}
Descriptive statistics (mean and standard deviation) of all previously mentioned evaluation criteria were summarized in Table~\ref{tab:descstats} shown below. 
\begin{table}[H]
\centering
\caption{Descriptive statistics (mean~$\pm$~SD) for pre-test, post-test, learning gain, engagement (IMI), and workload (NASA-TLX) across instructional conditions in the within-subject A-B crossover study ($N = 100$). All variables reflect paired observations from the same participants under both animation and slide-based formats.}
\label{tab:descstats}
\begin{tabular}{lcccc}
\hline
\textbf{Variable} & \textbf{Condition} & \textbf{Mean} & \textbf{SD} & \textbf{N} \\
\hline

\multirow{2}{*}{Pre-Test Score (0-100)} 
  & Animation                           & 60.51 & 9.54 & 100 \\
  & \cellcolor{lightblue!30}Slides      & \cellcolor{lightblue!30}59.47 
                                        & \cellcolor{lightblue!30}9.03 
                                        & \cellcolor{lightblue!30}100 \\[2pt]

\multirow{2}{*}{Post-Test Score (0-100)} 
  & Animation                           & 74.42 & 9.56 & 100 \\
  & \cellcolor{lightblue!30}Slides      & \cellcolor{lightblue!30}69.05 
                                        & \cellcolor{lightblue!30}9.38 
                                        & \cellcolor{lightblue!30}100 \\[2pt]

\multirow{2}{*}{Learning Gain (Post-Pre)} 
  & Animation                           & 13.91 & 4.90 & 100 \\
  & \cellcolor{lightblue!30}Slides      & \cellcolor{lightblue!30}9.58  
                                        & \cellcolor{lightblue!30}5.51 
                                        & \cellcolor{lightblue!30}100 \\[2pt]

\multirow{2}{*}{IMI (Engagement 1-7)} 
  & Animation                           & 5.43 & 0.44 & 100 \\
  & \cellcolor{lightblue!30}Slides      & \cellcolor{lightblue!30}4.89 
                                        & \cellcolor{lightblue!30}0.42 
                                        & \cellcolor{lightblue!30}100 \\[2pt]

\multirow{2}{*}{NASA-TLX (Workload 0-20)} 
  & Animation                           & 9.99 & 1.56 & 100 \\
  & \cellcolor{lightblue!30}Slides      & \cellcolor{lightblue!30}10.73 
                                        & \cellcolor{lightblue!30}1.21 
                                        & \cellcolor{lightblue!30}100 \\[2pt]

\hline
\end{tabular}
\vspace{0.8em}
\scriptsize{\textit{Note:} Values based on within-subject A-B crossover data ($N=100$). 
IMI = Intrinsic Motivation Inventory; NASA-TLX = NASA Task Load Index.}
\end{table}
We analyze these data, together with additional quantitative evidence collected through surveys, in the following subsections.

\subsection{Learning Outcomes}
Descriptive statistics for learning performance under both instructional conditions are summarized in Table~\ref{sec:results}. To address RQ1, we compared post-test performance while controlling for baseline differences and examined within-subject learning gains derived from the pre-post assessments.

\begin{table}[h]
\centering
\renewcommand{\arraystretch}{1.15}
\caption{Inferential results for learning, engagement, workload, satisfaction, and efficiency (N = 100).}
\label{tab:inferential-results}
\begin{tabular}{lccc}
\hline
\textbf{Measure} & \textbf{$t$(99)} & \textbf{$p$} & \textbf{Cohen's $d$} \\
\hline
Learning Gain & 6.74 & $<.001$ & 0.67 \\
IMI (Engagement) & 9.44 & $<.001$ & 0.94 \\
NASA-TLX (Workload) & -4.06 & $<.001$ & 0.41 \\
Satisfaction & 16.35 & $<.001$ & 1.64 \\
Efficiency (Time) & -8.56 & $<.001$ & 0.86 \\
\hline
\end{tabular}
\end{table}
Post-test performance was analyzed using analysis of covariance, a statistical method that adjusts group comparisons by controlling for baseline differences through a covariate (e.g., pre-test score). In subsequent text, this method is referred to as ANCOVA. The results of ANCOVA revealed a significant effect of instructional condition, $F(1,197)=38.85$, $p<.001$, with a partial $\eta^2=0.165$. Students performed substantially better after completing the animation-based module than after the slides-based module, with adjusted post-test means of $M_{\text{adj}}=83.4$ (Animation) and $M_{\text{adj}}=78.1$ (Slides). Assumption checks indicated no violations of normality, homogeneity of variance, or homogeneity of regression slopes, validating the use of ANCOVA for adjusted comparisons.

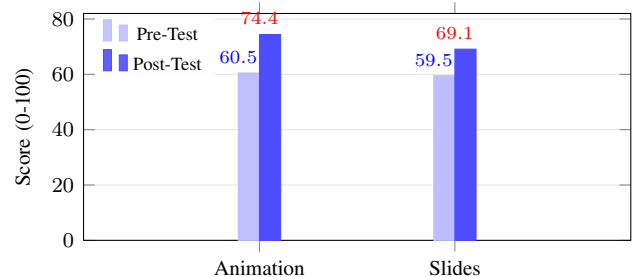
\begin{figure}[H]
\centering

\begin{tikzpicture}
\begin{axis}[
  width=\columnwidth,
  height=4.6cm,
  ybar,
  ymin=0, ymax=82,
  title style={font=\footnotesize},
  ylabel={Score (0-100)},
  ylabel style={font=\footnotesize},
  symbolic x coords={Animation,Slides},
  xtick=data,
  xticklabel style={font=\footnotesize},
  tick label style={font=\footnotesize},
  enlarge x limits=0.90,
  bar width=8pt,
  /pgfplots/bar shift=-4pt,
  ymajorgrids=true,
  grid style={gray!20},
  clip=false,
  nodes near coords,
  nodes near coords style={
    font=\scriptsize,
    /pgf/number format/fixed,
    /pgf/number format/precision=1,
    yshift=0.3pt,
    xshift=-4pt
  },
  legend style={
    font=\scriptsize,
    draw=none,
    fill=white,
    fill opacity=0.9,
    text opacity=1,
    at={(0.02,1)},
    anchor=north west
  },
]

\addplot+[fill=blue!25, draw=blue!25] coordinates {(Animation,60.5) (Slides,59.5)};
\addlegendentry{Pre-Test}

\addplot+[fill=blue!70, draw=blue!70, /pgfplots/bar shift=4pt]
coordinates {(Animation,74.4) (Slides,69.1)};
\addlegendentry{Post-Test}

\end{axis}
\end{tikzpicture}

%

\caption{Learning performance across instructional conditions, shown through the mean pre- and post-test scores for animation and slide-based instructions.}
\label{fig:learningcombined}
\end{figure}

To further evaluate learning improvement, paired-samples $t$-tests were conducted on learning gains. Students achieved significantly larger gains after the animation module ($M=13.91$, $SD=4.90$) relative to the slides module ($M=9.58$, $SD=5.51$), $t(99)=6.74$, $p<.001$, $d=0.67$. This effect size represents a medium-to-large within-subject impact, indicating that animation-based explanations were supported with greater conceptual improvement. Learning differences were also robust across instructional order: Animation-First and Slides-First sequences did not differ significantly in gain patterns ($p=.808$).

A consolidated visualization of learning performance is provided in Fig.~\ref{fig:learningcombined}. It displays the mean pre- and post-test scores for both instructional formats, and the right panel presents mean learning gains.
Students achieved significantly higher adjusted post-test scores and larger learning gains in the animation condition than in the slides condition, as reported in Table~\ref{tab:inferential-results}. 

\subsection{Engagement (IMI)}
Student engagement was measured using the six-item Intrinsic Motivation Inventory (IMI), which captures learners’ perceived enjoyment, interest, and value during each instructional module. Participants reported consistently higher engagement when interacting with the animation-based materials compared to the slide-based version. Mean IMI ratings were significantly higher for the animation condition ($M = 5.43$, $SD = 0.44$) than for the slides condition ($M = 4.89$, $SD = 0.42$), $t(99)=9.44$, $p<.001$, $d=0.94$. This represents a large within-subject effect, indicating that the animated explanations were perceived as more enjoyable and motivating.

Reliability analysis showed strong internal consistency for the IMI scale ($\alpha = .82$, 95\% CI [.78, .86]), confirming that the scale adequately captured engagement across conditions. IMI item distributions showed a consistent upward shift for the animation condition. Engagement differences were also robust across instructional order, with no significant sequence effects observed.

\begin{figure}[ht]
    \centering
    \includegraphics[width=0.48\textwidth]{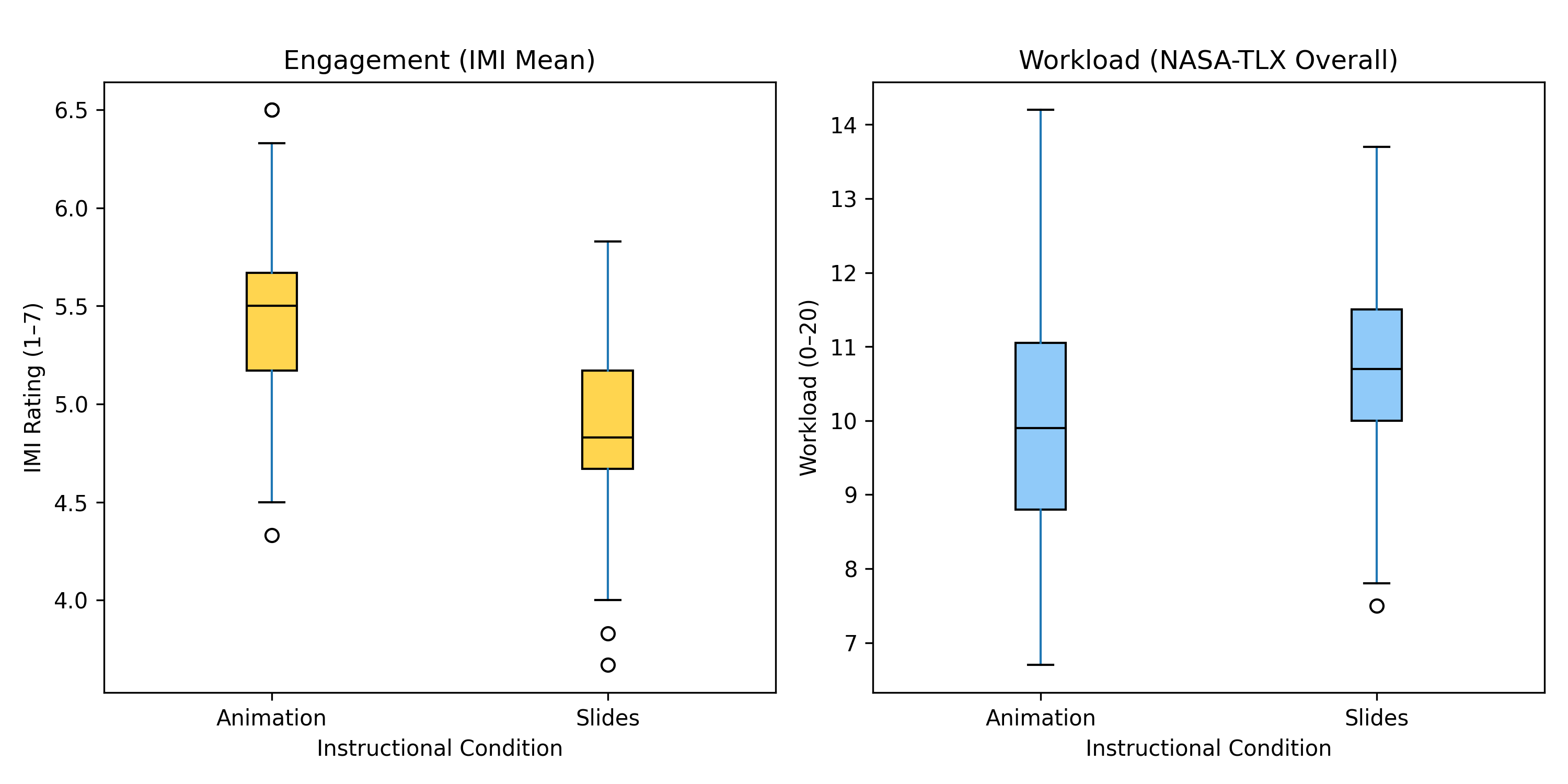}
    \caption{Engagement and cognitive workload across instructional conditions. 
    The left panel shows IMI engagement ratings for animation and slide-based modules. 
    The right panel presents overall NASA-TLX workload scores. 
    Boxplots indicate median, interquartile range, and outliers.}
    \label{fig:engageworkload}
\end{figure}

These findings align with multimedia learning theory, which predicts that coordinated narration and visual cues enhance learner interest and motivational investment. A visual summary of IMI ratings is provided in Fig.~\ref{fig:engageworkload} (left panel).
consistently higher IMI ratings for animations is shown, indicating that synchronized narration and directed visual attention cues increased learners’ interest and perceived value. At the same time, overall workload decreased (shown on the right panel), suggesting that the animations reduced extraneous cognitive load by clarifying symbolic and spatial relationships, consistent with cognitive load theory.

\subsection{Cognitive Workload (NASA-TLX)}
Cognitive workload was assessed using the NASA-TLX instrument, which captures perceived mental, physical, temporal, and effort-related demands of the learning task. Students reported lower workload during the animation module ($M = 9.99$, $SD = 1.56$) than during the slides module ($M = 10.73$, $SD = 1.21$). A paired-samples $t$-test confirmed this difference, $t(99)=-4.06$, $p<.001$, corresponding to a moderate effect size ($d = 0.41$, 95\% CI [0.21, 0.61]). This reflects a moderate reduction in perceived workload for the animation condition.

Reliability analysis showed acceptable internal consistency for the NASA-TLX scale ($\alpha = .79$). Workload differences were also robust across instructional order, with no significant sequence effects observed. These findings align with cognitive load theory, suggesting that synchronized narration and targeted signaling reduced extraneous load by helping learners focus on the most relevant symbolic or spatial transformations.

A visual summary of workload ratings is presented in Fig.~\ref{fig:engageworkload} (right panel), illustrating consistent downward shifts across TLX subscales for the animation condition.

\subsection{Satisfaction}

To address RQ5, overall learner satisfaction with each instructional medium was compared using paired-samples $t$-tests. Participants reported substantially higher satisfaction with the animation-based modules ($M = 5.63$, $SD = 0.41$) compared to the slide-based modules ($M = 4.79$, $SD = 0.47$). This difference was statistically significant, $t(99) = 16.35$, $p < .001$, with a very large effect size ($d = 1.64$, 95\% CI [1.32, 1.96]). These results directly address RQ5 and indicate a strong overall preference for the animation condition.

The magnitude of this effect is consistent with the system’s design philosophy, synchronized narration, segmentation, and signaling, which collectively enhance clarity and reduce perceived effort during learning. High satisfaction ratings, therefore, complement the observed gains in engagement and learning performance, suggesting that learners found the AI-generated animations both effective and enjoyable. A summary of satisfaction outcomes is presented in Table~\ref{tab:inferential-results}.

\subsection{Efficiency}

To address RQ4, exploratory analyses were conducted to compare task completion efficiency across instructional conditions. Although precise end-times were not logged during deployment, coarse timing metadata from survey submissions and learner self-reports enabled approximate duration estimates for each module. Participants completed the animation-based module more quickly ($M = 11.25$ minutes, $SD = 1.18$) than the slides-based module ($M = 13.07$ minutes, $SD = 1.31$). A paired-samples $t$-test confirmed this difference, $t(99) = -8.56$, $p < .001$, corresponding to a large within-subject effect ($d = -0.86$). These results directly address RQ4 and indicate that the animation condition supported more efficient completion of concept-focused tasks.

Because timing granularity was limited, these findings are interpreted cautiously; however, when considered alongside reduced NASA-TLX workload ratings, they suggest that the animation condition lowered extraneous cognitive load and facilitated more rapid information processing. Efficiency differences were also robust across instructional order, with no significant sequence effects observed. A summary of inferential outcomes, including efficiency results, is provided in Table~\ref{tab:inferential-results}.

\subsection{Equity and Subgroup Effects}

To address RQ3, we examined whether the advantage of animation-based instruction varied across learner subgroups, particularly those with lower prior preparation. Participants were grouped into Low (None + Basic) and High (Intermediate + Advanced) prior-knowledge categories. Analysis of participant-level learning gain differences (Animation - Slides) revealed a statistically significant subgroup effect: learners in the Low prior-knowledge group showed a smaller gain from slide-based instruction and therefore exhibited a larger benefit from animations ($M = 2.31$) compared to their High prior-knowledge peers ($M = 5.15$), $t(98) = -2.11$, $p = .040$. This pattern suggests that animation-supported explanations provided additional scaffolding that narrowed performance disparities for lower-prepared learners, consistent with Hypothesis H3.

Robustness checks confirmed that this effect was not attributable to sequencing or topic differences. No significant order effects were observed between Animation-First and Slides-First groups ($p = .808$), and condition-by-topic and condition-by-period interactions were non-significant. These findings indicate that the observed animation advantage generalized across instructional sequences, topics, and module positions.

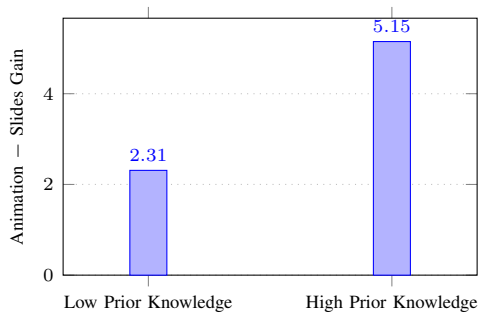
\begin{figure}[t]
\centering
\begin{tikzpicture}
\begin{axis}[
    ybar,
    ymin=0,
    width=0.8\linewidth,
    height=5cm,
    bar width=14pt,
    enlarge x limits=0.35,
    ylabel={Animation $-$ Slides Gain},
    symbolic x coords={Low,High},
    xtick=data,
    xticklabels={Low Prior Knowledge, High Prior Knowledge},
    ymajorgrids=true,
    grid style={dotted},
    ticklabel style={font=\scriptsize},
    label style={font=\scriptsize},
    nodes near coords,
    nodes near coords align={vertical},
    every node near coord/.append style={font=\scriptsize, anchor=south},
    every axis plot/.style={fill=gray!35}
]

\addplot coordinates {
    (Low, 2.31)
    (High, 5.15)
};

\end{axis}
\end{tikzpicture}

\caption{
Animation advantage in learning gains (Animation - Slides) by prior-knowledge subgroup. 
Both groups benefited from animations, with a larger gain difference observed for learners 
with higher prior knowledge.
}
\label{fig:equity-subgroup}
\end{figure}

Figure~\ref{fig:equity-subgroup} visualizes these subgroup differences, highlighting the disproportionate benefit for learners with lower initial preparation. Together, these results support the equity potential of pedagogy-guided, narration-synchronized animations in STEM education.
Importantly, these results reveal that although both groups benefited from the animations, higher-prepared learners exhibited larger animation advantages, indicating that prior conceptual scaffolding may have enabled deeper uptake of the narrated visual explanations. Together, these results validate Hypotheses H1 and H2 and confirm that narration-synchronized, pedagogically structured animations can improve learning outcomes and subjective experience without increasing workload.

\subsection{Qualitative Feedback}

Open-ended survey responses were analyzed thematically to contextualize the quantitative findings. Students provided brief comments after each module describing what helped or hindered their learning. Two researchers independently reviewed the responses, iteratively grouped similar statements into themes, and resolved disagreements through discussion. The resulting themes offer convergent evidence for the benefits and trade-offs of animation-based instruction.

A first theme, \emph{Enhanced Engagement and Enjoyment}, reflected frequent descriptions of the animations as “more interesting,” “fun,” or “attention-grabbing.” Learners reported that motion and synchronized narration “kept me focused” and reduced the temptation to multitask compared to slides. These comments align with the large IMI effect favoring the animation condition.

A second theme, \emph{Conceptual Clarity}, highlighted that animations made abstract mathematics and physics “easier to visualize” and “showed what the symbols were doing.” Students remarked that seeing transformations unfold step-by-step helped them connect formulas to geometric or physical interpretations, reinforcing the observed gains in post-test performance.

A third theme, \emph{Pacing and Control}, captured mixed preferences. Several participants appreciated the fixed pacing of animations for “keeping me moving” through the material, whereas others preferred slides for pausing and re-reading at their own speed. This tension suggests that adding lightweight controls (pause, replay, seek) could further improve the usability of animation-based materials.

A fourth theme, \emph{Cognitive Load and Focus}, echoed the workload results. Many students indicated that animations “highlighted what to pay attention to” and reduced the need to infer intermediate steps, while a smaller number reported occasional overload when multiple visual elements moved simultaneously. Overall, the qualitative patterns support the quantitative evidence that pedagogy-guided animations increase engagement and understanding while generally maintaining manageable cognitive load.

\subsection{Summary of Inferential Results}

Table~\ref{tab:inferential-results} consolidates all inferential outcomes across learning, engagement, workload, satisfaction, and efficiency. Across all measures, the animation condition yielded statistically significant improvements with medium-to-large effect sizes. Learning gains showed a medium-to-large impact ($d = 0.67$), engagement demonstrated a large effect ($d = 0.94$), workload showed a moderate reduction ($d = 0.41$), and satisfaction exhibited an exceptionally large effect ($d = 1.64$). Efficiency analyses further indicated that learners completed the animation module more quickly, with a large efficiency effect ($d = 0.86$). Together, these convergent results highlight the robust pedagogical and experiential advantages of narration-synchronized, pedagogy-guided Manim animations.

\section{Discussion}
\label{sec:discussion}

\subsection{Implications for Learning Technology Design and Practice}

While the results demonstrate that LLM-generated, pedagogy-aware animations can meaningfully enhance learning and learner experience in STEM contexts, they also have several implications for instructional practice and the integration of AI-generated media in STEM education. Learners not only achieved higher post-test performance with the animation condition (Fig.~\ref{fig:learningcombined}) but also reported substantially greater engagement and enjoyment (Fig.~\ref{fig:engageworkload}, left panel). Combined with the moderate reduction in perceived workload (Fig.~\ref{fig:engageworkload}, right panel), these findings suggest that pedagogy-informed animations can both support deeper understanding and create a more motivating learning environment. This pattern aligns with cognitive load theory and multimedia learning research \cite{mayer2014cambridge, sweller2011cognitive}, which predicts that segmentation, signaling, and synchronized narration reduce extraneous processing demands and promote dual-channel integration.

The very large satisfaction effect ($d = 1.64$) and the sizable efficiency advantage ($d = 0.86$), summarized in Table~\ref{tab:inferential-results}, highlight the practical benefits of the medium. These outcomes indicate that students not only learned more but did so with less perceived effort and in less time, reinforcing the usability of the generated animations in both classroom and self-paced learning environments.

For instructors, the LLM-Manim pipeline reduces the expertise and time required to create high-quality visual explanations, enabling wider adoption of animations beyond experts with specialized programming or design backgrounds. This makes it feasible to incorporate concise, concept-focused animations into blended learning models, flipped classroom structures, and online instructional modules. The positive reception observed in this study suggests strong learner readiness for such media, and the workflow’s reproducibility and HITL quality control support instructional reliability across offerings.

\subsection{Equity, Limitations and Future Directions}
To address RQ3, we examined whether the advantages of animation-based instruction varied across learner subgroups. As shown in Fig.~\ref{fig:equity-subgroup}, both lower and higher prepared learners benefited from the animation condition, although higher-prepared learners exhibited a larger animation-related gain. This pattern suggests that prior conceptual scaffolding enabled deeper uptake of the narrated visual explanations, while lower-prepared learners still achieved meaningful improvements. These findings highlight that animation-based explanations can support a broad range of learners but may require additional scaffolds such as slower pacing, selective replay, or introductory warm-up prompts to maximize benefits for novices.

Beyond prior knowledge differences, the system’s built-in accessibility features, such as automatic captioning, consistent narration, controlled pacing, and color stable design, support learners with varied processing preferences, including non-native speakers and students with attention or working memory challenges. Although accessibility was not a primary focus of the present evaluation, these affordances align with universal design principles and suggest that LLM-generated animations may serve as an inclusive medium for diverse STEM learning needs.

Future equity-focused work should expand these analyses across demographic groups such as gender, linguistic background, and socioeconomic status, and examine whether adaptive scaffolding or personalized pacing can narrow remaining performance gaps.

This study was conducted with two instructors at a single institution, which may limit generalizability. Efficiency measures relied on coarse timing data, and satisfaction was assessed using a single-item scale. Learning outcomes were measured immediately after instruction, leaving long-term retention unexamined. The pipeline also requires human-in-the-loop review, and LLM behavior may vary across model updates. Finally, only two STEM topics were included, suggesting the need for broader curricular evaluation.

Future work should examine long-term retention and larger, multi-institution deployments to assess the durability and generalizability of the observed effects. Additional research is needed to extend the pipeline to a broader range of STEM topics, reduce reliance on human-in-the-loop review, and improve robustness across LLM model updates. A promising direction involves transitioning from a single-model workflow to a multi-agent pipeline, in which specialized agents collaboratively critique, refine, and validate narration, code, and pedagogical structure to further enhance automation accuracy and instructional quality. Equity-focused evaluations spanning gender, linguistic background, and socioeconomic status are also essential, along with targeted scaffolds to further support lower-prepared learners. Finally, adding adaptive pacing, multilingual narration, and accessibility features such as richer captioning and interactive controls represents a promising direction for expanding the instructional reach of AI-generated animations.

\section{Conclusion}
\label{sec:conclusion}
This work introduced an LLM-driven, pedagogy-aware pipeline for generating narrated STEM learning animations using the Manim framework. By embedding multimedia learning principles into structured prompt templates, parallel narration–code generation, and a three-stage human-in-the-loop review, the system transforms natural-language concept descriptions into synchronized visual verbal explanations with high instructional fidelity.

A controlled classroom study with 100 undergraduate STEM learners demonstrated that these AI-generated animations yield meaningful educational benefits. Across topics, students achieved higher post-test performance, reported greater engagement, and experienced lower cognitive workload compared to traditional slide-based instruction. Satisfaction and efficiency advantages further indicated that learners not only understood more, but did so with less perceived effort and in less time. Subgroup analyses showed that the animation benefit generalized across learners with different levels of prior knowledge, with higher-prepared students exhibiting the largest gains. Qualitative feedback echoed these findings, highlighting improvements in clarity, motivation, and conceptual visualization.

Together, these results provide the first controlled evidence that LLM-generated, pedagogy-guided animations can support effective STEM learning at scale. Beyond demonstrating feasibility, the current study establishes a foundation for automated instructional media that is both technically robust and grounded in learning science. As LLM capabilities continue to evolve, future extensions of this framework, including adaptive pacing, personalized scaffolding, multilingual narration, and richer accessibility support offer substantial potential for democratizing the production of high-quality, cognitively aligned STEM explanations.

\section*{Acknowledgments}
This work was supported by the California Learning Lab under the AI Fast-Funding for Accelerated  Study and Transformation, through the project ``Animate Math Concepts in Engineering Education using Large Language Models.”
\bibliographystyle{IEEEtran}
\bibliography{references}
\nocite{*}

\end{document}